\documentclass[letterpaper, 10 pt, conference]{ieeeconf}  
\usepackage{graphicx}
\usepackage{epstopdf}
\usepackage{color}
\UseRawInputEncoding
\IEEEoverridecommandlockouts                              

\title{\LARGE \bf
Users' Perception on Appropriateness of Robotic Coaching Assistant's Disclosure Behaviors 
}

\author{Atikkhan Faridkhan Nilgar$^{1}$, Manuel Dietrich$^{2}$ and Kristof Van Laerhoven$^{1}$
\thanks{This work was supported by Honda Research Institute Europe GmbH.}
\thanks{$^{1}$Atikkhan Faridkhan Nilgar is with Ubiquitous Computing, University of Siegen, Hölderlinstrasse 3, 57076 Siegen, Germany
        {\tt\small atikkhan.nilgar@uni-siegen.de}}%
\thanks{$^{1}$Kristof Van Laerhoven is with Ubiquitous Computing, University of Siegen, Hölderlinstrasse 3, 57076 Siegen, Germany
        {\tt\small kvl@eti.uni-siegen.de}}%
\thanks{$^{2}$Manuel Dietrich is with Honda Research Institute Europe GmbH, 
        Carl-Legien-Straße 30, 63073 Offenbach am Main, Germany
        {\tt\small manuel.dietrich@honda-ri.de}}%
}

\begin{document}

\maketitle
\thispagestyle{empty}
\pagestyle{empty}

\begin{abstract}

Social robots have emerged as valuable contributors to individuals' well-being coaching. Notably, their integration into long-term human coaching trials shows particular promise, emphasizing a complementary role alongside human coaches rather than outright replacement. In this context, robots serve as supportive entities during coaching sessions, offering insights based on their knowledge about the users' well-being and activity. Traditionally, such insights have been gathered through methods like written self-reports or wearable data visualizations. However, the disclosure of people’s information by a robot raises concerns regarding privacy, appropriateness, and trust. To address this, we conducted an initial study [\textit{n} = 22] to quantify participants' perceptions of privacy regarding disclosures made by robot coaching assistant. The study was conducted online, presenting participants with six prerecorded scenarios illustrating various types of information disclosure and the robot's role, ranging from active on-demand to proactive communication conditions.

\end{abstract}

\section{INTRODUCTION}

Social robots have shown significant potential in the field of well-being coaching. In human-robot interaction (HRI), robotic coaches have been examined in different environments such as workplace \cite{Micol2023RM, Spitale2023VITAAM}, public \cite{Minja2023RC, Kayla2022ASocial}, lab \cite{Minja2022ParticipantPO, Bodala2021Tele,Churamani2022Continual} and at home \cite{Jeong2020ARP, Jeong2023ARC, Jeong2022DRP}. In most approaches the robot takes the role of coach for instance through everyday well-being interventions or reinforcing positive habits \cite{Matous2024DevAuto}.

Social robots are investigated less as coaching assistants in human-human coaching sessions. For instance, the information a social robot has acquired about individuals’ (their owner) habits, and emotional level can be very useful to be shared with a professional coach and taken forward by them. Such a supportive information providing role is promising, since we know of the need for well-being and activity brought into sessions from traditional therapy. That is why methods like written self-reports or even advanced techniques like wearable data analysis are used \cite{Laban2023, Leite2013SocialRF}. However, robot disclosure of personal information in front of others, even to a human coach does raise valid concerns regarding privacy and appropriateness \cite{Lutz2021DoPrivacy, Yang2021EffectsOS}.

In our approach, we looked at scenarios where a social robot serves as a supportive entity during coaching sessions, providing insights based on their understanding of the users' well-being and activity. Complementary to this, robots could also directly receive this data from users when they choose to share it from on-body wearables and sensors in personal devices \cite{Bozcan}. 

In this preliminary study, we aimed to measure how participants perceived privacy concerning the information disclosed by robot coaching assistants. Therefore, we seek to determine what impact the proactive role of the robot has on users' perception. We know from prior work that a robot’s degree of proactivity and independence significantly influences acceptance of robot behavior and the resultant HRI experience \cite{chien2022theeffect}, \cite{Buyukgoz2021Explor}. However, independent decision-making power is also a potential reason for human concerns. The study was carried out as an online survey, where participants were exposed to six prerecorded scenarios. These scenarios were carefully designed to represent a variety of information disclosure types and the role of the robot, which varied from providing information as active on-demand and proactive conditions. 

We aim to address the following research questions:

\begin{itemize}
\item \textbf{RQ1:} Does the proactivity of a robotic coaching assistant influence people's judgments of disclosure appropriateness?

\item \textbf{RQ2:} Does the type of information shared by a robotic coaching assistant influence people's judgments of disclosure appropriateness?
\end{itemize}

This study is part of our ongoing efforts to understand the HRI in the context of disclosure appropriateness, particularly focusing on the delicate balance between providing coaching assistance and maintaining users' privacy. The insights gained from this study will inform the design and development of future robot coaching assistants, ensuring they are both effective in their role and respectful of disclosing users' private information. Furthermore, ensuring that social robots respect users' privacy while providing valuable assistance and support is essential for fostering trust and acceptance among users.

\section{RELATED WORK}

Prior work on the appropriateness of robot disclosure examines how suitable and acceptable it is for robots to share information in a given context such as in domestic environments \cite{ dietrich2022, cagiltay2023designing}. A research examined suitable strategies for disclosure by robots \cite{dietrich2023what}. Participants received written scenarios illustrating social robots disclosing personal details about their owners to enhance human-human interaction. They were asked to choose from various robot behaviors differing in the degree of information disclosure. The results of this study on information content, relationship configurations, privacy expectations, and privacy attitudes provide insights into how robots can disclose private information in a manner that users perceive as appropriate.

A research on people’s attitudes about a service robot using customer data in conversation \cite{Hedaoo2019RoboBarista} was conducted in a cafe setting, which is considered a social grey area, mixing both public and private spaces. During the experiment, participants were subjected to privacy violations using the Theater Method rather than experiencing actual breaches of their privacy. The research examined various factors influenced by social dynamics related to data, including the origin of data, its beneficial or detrimental utilization, and the audience targeted by the robot's verbal communication \cite{Hedaoo2019RoboBarista}. All of these factors were found to be potent predictors of participants' social attitudes regarding the robot's appropriateness \cite{Hedaoo2019RoboBarista}.

Furthermore, a privacy controller called CONFIDANT \cite{tang2022confidant} was designed. This controller utilizes contextual metadata, such as sentiment, relationships, and topic of conversations to model privacy boundaries. The researchers first assessed human-human interaction scenarios for privacy sensitivity and then evaluated the effectiveness of the privacy controller in HRI. The study suggests that the privacy controller can help manage the information shared with the robots, discerning the sensitivity of the information, which is crucial for human-robot trust. The controller can handle disclosures in a way that respects the privacy of the users \cite{tang2022confidant}.

Prior studies have examined how robots disclose information in privacy settings \cite{dietrich2023what, Hedaoo2019RoboBarista, tang2022confidant}, but they have not explored user perceptions of disclosures when robots serve as coaching assistants. Moreover, previous research hasn't delved into how users perceive the appropriateness of disclosures from proactive communication of robots.

\section{STUDY DESIGN}

\subsection{Participants}

We sent a survey link via the university student email list as well as a social media channel (LinkedIn), inviting participation in our study. 22 participants completed the entire survey, with 16 identified as male, 5 as female, and 1 as diverse. 86\% of participants fell within the age range of 18 to 49 years. 14 participants held a university degree, while 8 participants held a doctoral degree. 8 participants have prior experience with social robots and 14 participants have used smartwatches capable of tracking health activities.

\subsection{System and Scenario}

\begin{figure}[h]
  \centering
  \includegraphics[width=1\linewidth]{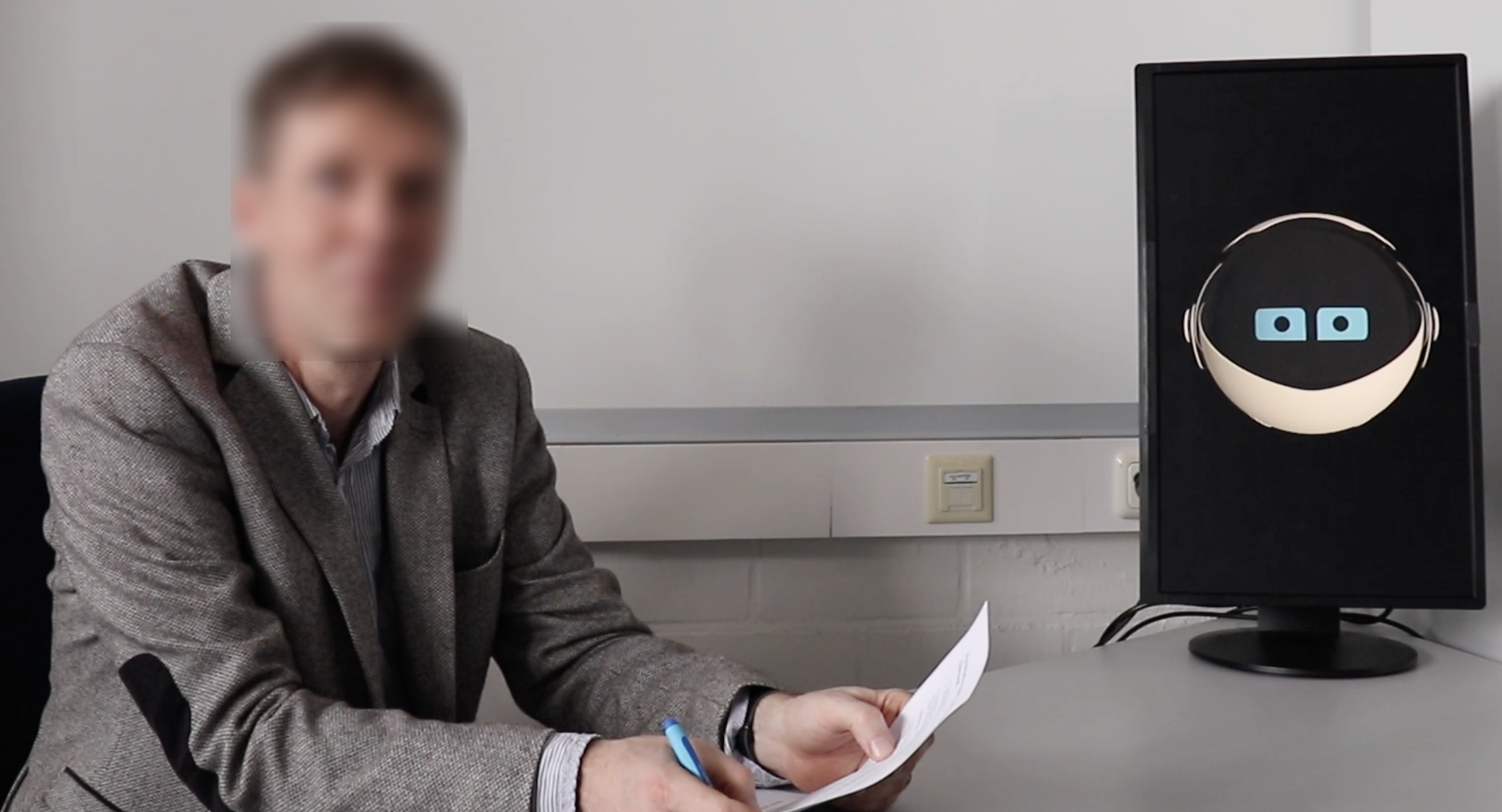}
  \caption{Participants were shown short first-person-perspective video clips in which a coach and the personal assistant Meisy discuss about the data from the participant's smartwatch}
  \label{CoachandMeisy}
\end{figure}

\begin{figure}[h]
  \centering
  \includegraphics[width=1\linewidth]{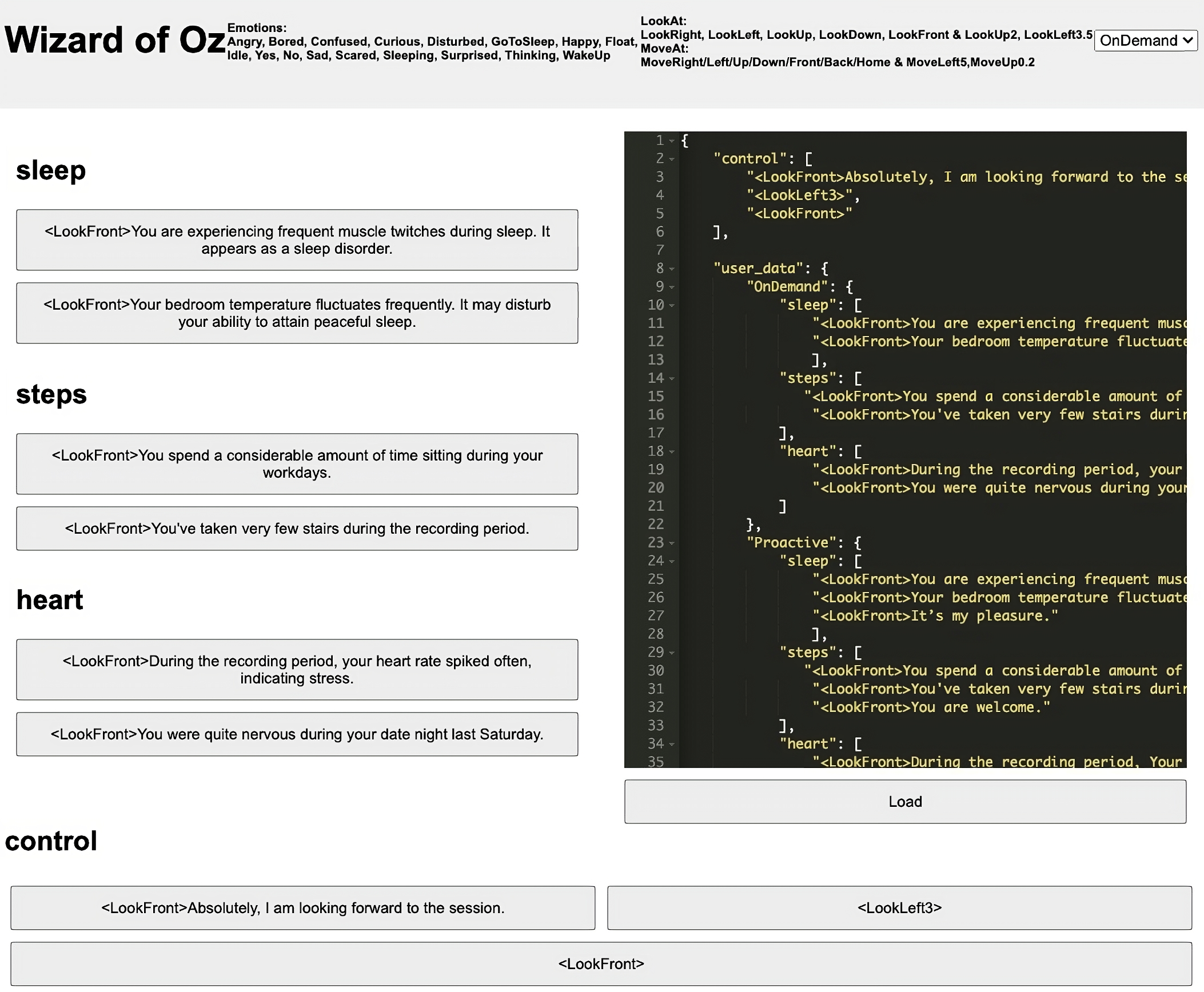}
  \caption{GUI to command Meisy dialogues with text-to-speech and change the gaze of Meisy (left and right)}
  \label{WOZ}
\end{figure}

We employed a 3D virtual avatar robot named ‘Meisy’ (Figure \ref{CoachandMeisy}), previously developed and deployed as a virtual receptionist \cite{wang2020designing}. We devised a graphical user interface (GUI) based platform (Figure \ref{WOZ}) for controlling Meisy's text-to-speech dialogues and movements from other computer as typically used in a Wizard of Oz study \cite{Kelley1984, Riek2012Wizard}. One of the involved researchers performed as a health and well-being coach, while Meisy served as the participant's personal assistant (Figure \ref{CoachandMeisy}). Two introduction videos were recorded: One featuring active on-demand communication condition and another featuring proactive communication condition. In the active on-demand condition, Meisy responded only when explicitly called upon or prompted by the coach. Proactive social robots anticipate users' needs and initiate interactions accordingly, engaging actively rather than merely reacting to commands \cite{Henschel2021whatmakes}. We recorded the proactive condition where Meisy responded without explicitly being called or prompted by the coach.

Smartwatches possess the capability to monitor various health indicators, including heart rate, sleep condition, and levels of physical activity \cite{Mohsen2023Smartwatches, Jat2022, alex2022}. We selected three types of activity from the smartwatch: Sleep tracking, step counting, and heart rate monitoring. We formulated and scripted these activities' information to ensure that they are perceived as private information to the participants. A total of six videos were recorded, three for each smartwatch activity, with both active on-demand and proactive conditions respectively. All videos featured both the coach and Meisy.

Table \ref{script_table} is an example script of step activity with active on-demand and proactive conditions. In the dialogues, the term “You” refers to the participant. Similarly, we created scripts for sleep and heart rate information.

\begin{table}[!ht]
    \centering
    \caption{Script dialogues on step information in active on-demand and proactive conditions}
    \label{script_table}
    \begin{tabular}{|l|p{7cm}|}
    \hline
    \multicolumn{2}{|c|}{\textbf{Active On-Demand}} \\ \hline
    \textbf{Coach} & Meisy, do you have any information on step count?\\ \hline
    \textbf{Meisy} & You spend a considerable amount of time sitting during your workdays. \\ \hline
    \multicolumn{2}{|c|}{\textbf{Proactive}} \\ \hline
    \textbf{Meisy} & (\textit{Meisy jump in the conversation of step count without being called}) You spend a considerable amount of time sitting during your workdays. \\ \hline
    \end{tabular}
\end{table}

\subsection{Procedure}

\begin{figure}[h]
  \centering
  \includegraphics[width=0.75\linewidth]{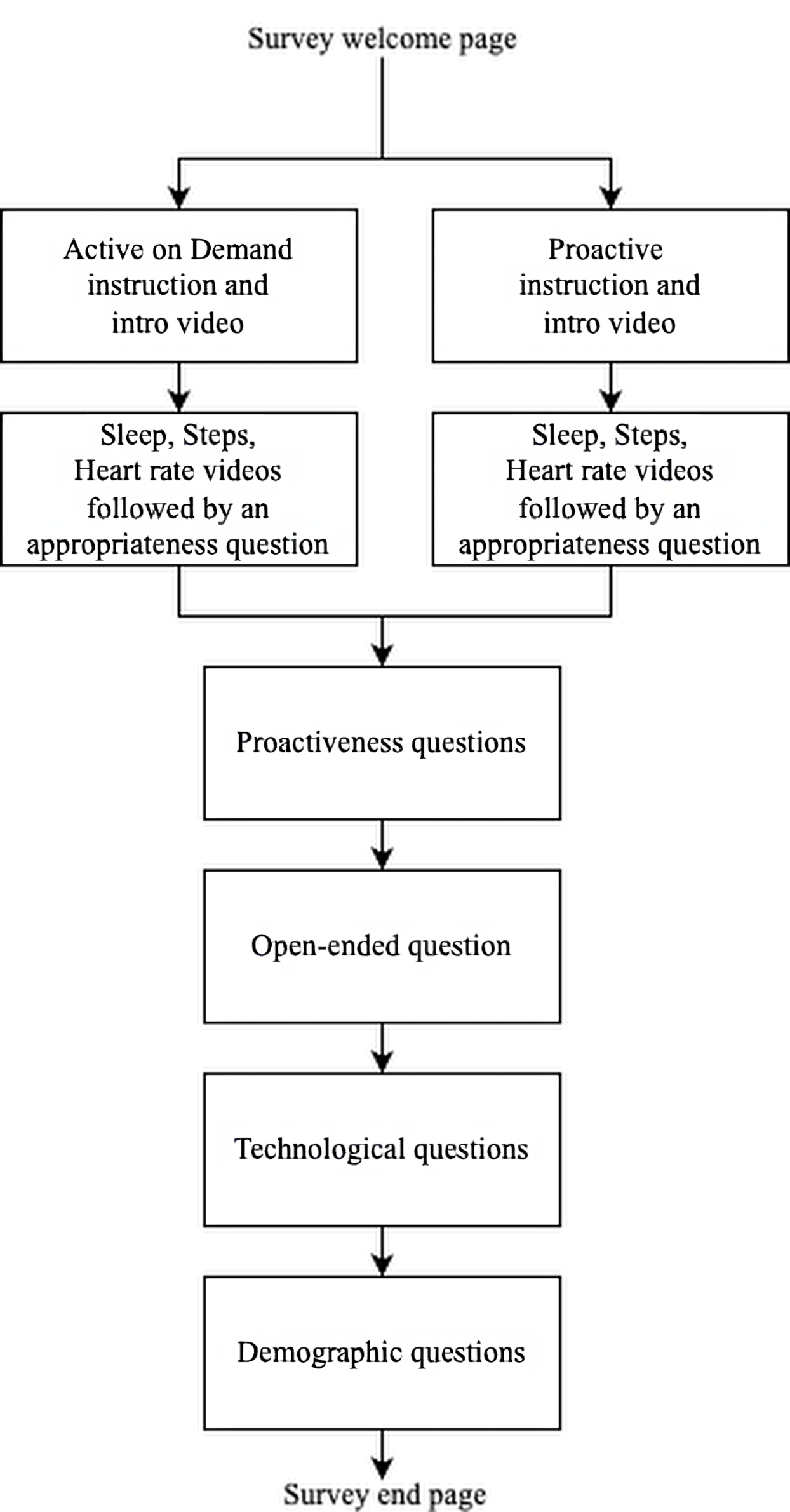}
  \caption{Flowchart of the survey procedure}
  \label{Flowofstudy}
\end{figure}

After participants accepted our invitation, the link directed them to LimeSurvey (https://www.limesurvey.org). Subsequently, each participant agreed to the terms and conditions. Participants were then randomly allocated to either the active on-demand or proactive conditions. Next, participants were instructed to imagine that Meisy functions as their personal assistive social robot with an access to their smartwatch data. Following this, participants viewed three videos according to their assigned conditions, with the order of videos also being randomized. Each video was displayed twice to ensure participants paid attention to all details. At the end of each video, participants were prompted to answer a question on disclosure behavior appropriateness, as outlined in \ref{Measures}, to evaluate their experience with Meisy. Additionally, participants from both conditions were asked to fill out a questionnaire assessing the perceived proactiveness of the personal assistant robot Meisy. Furthermore, an optional open question was asked. At the end, participants responded to a set of demographic questions and technical questions (prior experience with social robots and smartwatches). Figure \ref{Flowofstudy} represents the flowchart of the survey procedure.

\subsection{Measures}\label{Measures}

\subsubsection{Appropriateness}

We assessed the perceived appropriateness of Meisy's disclosure of information regarding sleep, steps, and heart rate. After each video, participants were asked to rate their agreement with the statement “Meisy's disclosure behavior was appropriate.” on a 5-point Likert scale ranging from “Strongly Disagree” (1) to “Strongly Agree” (5).

\subsubsection{Proactiveness}

To evaluate the perceived proactiveness of robot Meisy, we asked participants the following statements: “Meisy was aware of the surrounding environment.”, “Meisy was actively listening to the conversation.”, “Meisy initiated interactions without waiting for explicit coach requests.”, “Meisy adjusted its behavior based on context.” and “Meisy’s behavior was proactive.” Participants were asked to rate their agreement with the statements on a 5-point Likert scale ranging from “Strongly Disagree” (1) to “Strongly Agree” (5). The average of the responses to each element of the scale was calculated to generate a perceived proactiveness score.

\subsubsection{Open-Ended Question}

We asked an optional open-ended question to participants: “What do you like most about Meisy and what do you like least about Meisy?” The responses of participants were stored in a textual format. 

\subsection{Analysis}
We implemented means and standard deviation scores to analyze perceived disclosure appropriateness and perceived proactiveness. This statistical approach allowed us to assess the central tendency and variability of participants' responses. We organized the open-ended question responses based on their similarity and shared sentiments.

\section{RESULTS}

\begin{figure}[h]
  \centering
  \includegraphics[width=1\linewidth]{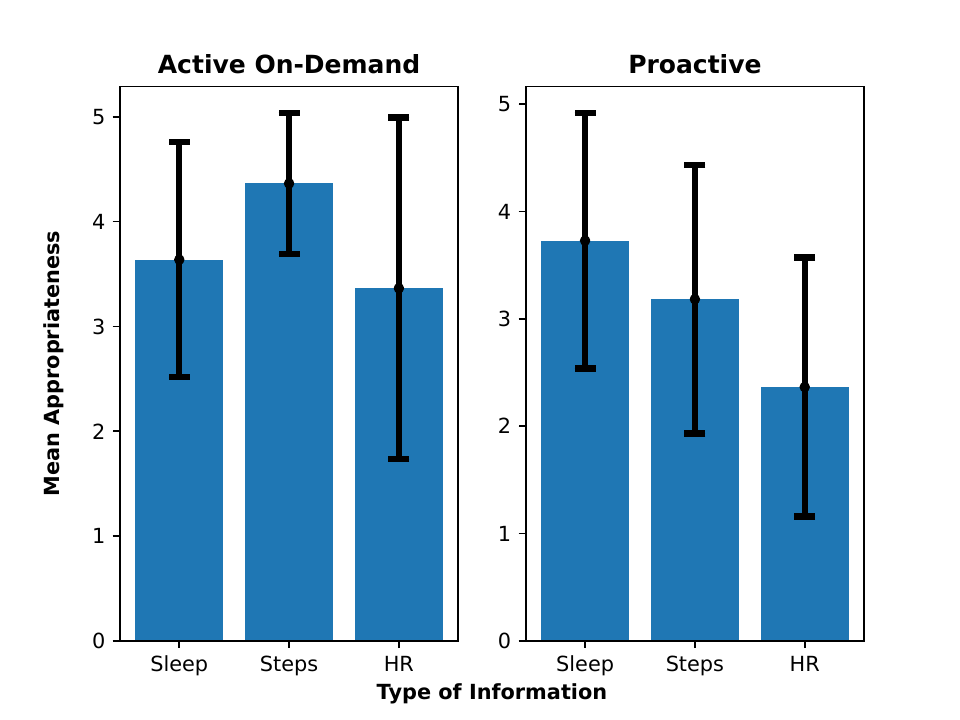}
  \caption{Disclosure appropriateness mean scores and standard deviations for active on-demand and proactive conditions with the activity type information (sleep, step and heart rate (HR))}
  \label{Appropriate}
\end{figure}

Among the 22 participants, half were randomly assigned to the active on-demand condition and the rest were assigned to the proactive condition. We identified mean differences in the appropriateness perception between the active on-demand and proactive conditions of sleep, steps and heart rate information disclosure. The active on-demand robot's steps information disclosure (\textit{M} = 4.36, \textit{SD} = 0.67) is perceived as more appropriate when compared to steps information disclosures by the proactive robot (\textit{M} = 3.18, \textit{SD} = 1.25). Similarly, heart rate information with the active on-demand robot (\textit{M} = 3.36, \textit{SD} = 1.62) is perceived as more appropriate when compared to heart rate information with a proactive robot (\textit{M} = 2.36, \textit{SD} = 1.20). Sleep information disclosed by the on-demand robot (\textit{M} = 3.63, \textit{SD} = 1.12) and sleep information with proactive robot (\textit{M} = 3.72, \textit{SD} = 1.19) show almost similar mean scores. Heart rate information is viewed as the least appropriate in both active on-demand and proactive robot conditions. All perceived disclosure appropriateness results can be found in Figure \ref{Appropriate}. 

\begin{figure}[h]
  \centering
  \includegraphics[width=1\linewidth]{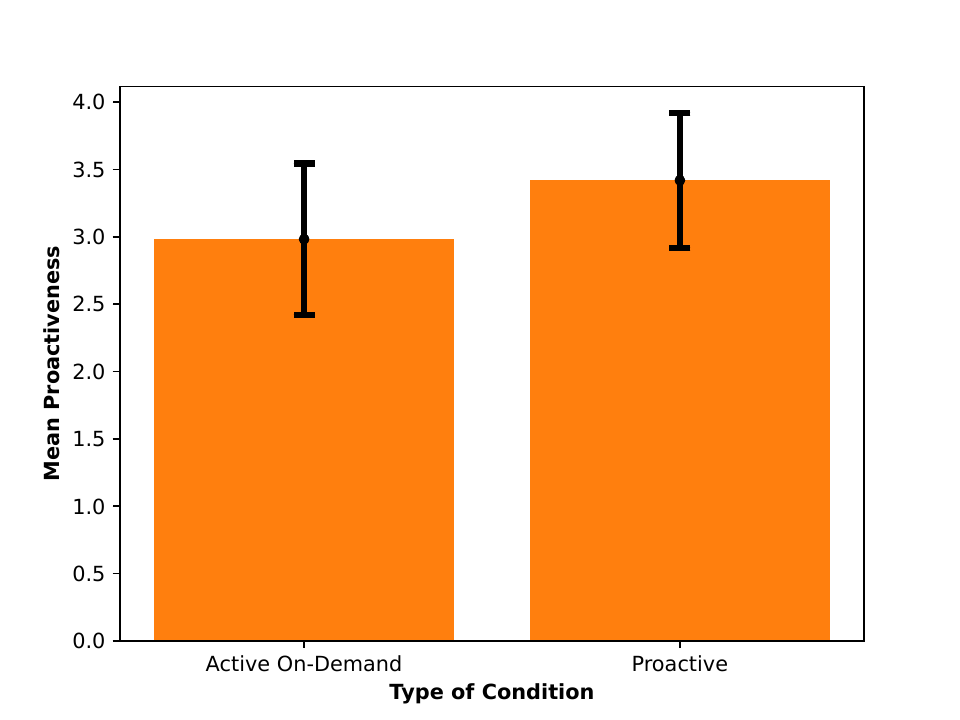}
  \caption{Perceived proactiveness mean scores and standard deviations for active on-demand and proactive conditions}
  \label{Proactiveness}
\end{figure}

The results showed that participants of the proactive condition (\textit{M} = 3.41, \textit{SD} = 0.50) perceived the robot as more proactive according to our scale compared to the active on-demand condition (\textit{M} = 2.98, \textit{SD} = 0.56). Means scores of the proactive perception scale can be found in Figure \ref{Proactiveness}.

Qualitative analysis of participants' experiences with Meisy through open-ended questions revealed varying perceptions based on both active on-demand and proactive conditions. Participants in the active on-demand condition expressed a desire for Meisy to be more proactive [\textit{n} = 2], while those in the proactive condition mentioned instances where Meisy interrupted the coach [\textit{n} = 2]. Moreover, some participants expressed strong privacy concerns, consistently indicating strong disagreement with the appropriateness of information disclosure and described Meisy as intrusive [\textit{n} = 3]. Their comments highlighted discomfort with Meisy disclosing private information and drawing medical conclusions from smartwatch data.

\section{DISCUSSION}

This study investigated users' perception on the appropriateness of robotic coaching assistant's disclosure in two communication conditions: active on-demand and proactive. Additionally, we used three types of activity information (sleep, step, and heart rate) of the smartwatch for each condition. 

Due to the limited number of participants, we cannot make a definitive judgment on the perceived appropriateness. The results of this preliminary study imply that the proactivity of a robotic coaching assistant might play a role in people's judgments of disclosure appropriateness, as addressed in \textbf{RQ1}. Additionally, the perceived appropriateness of steps and heart rate information indicate distinctions between the active on-demand and proactive conditions (see: \textbf{RQ2}).


The participants were exposed to prerecorded videos illustrating various types of information disclosure, these scenarios may not encompass the full range of potential situations in real-life coaching scenarios. The smartwatch data information in our study did not include real private data from individuals similar to other disclosure appropriateness studies \cite{Martin2012, dietrich2023what}. However, we acknowledge that future studies including a larger number of participants and implementing real data in an offline study might produce concrete results. Additionally, employing additional statistical analysis (i.e., independent-samples t-test, Mann-Whitney U test) in future investigations could offer further insights into this research. Moreover, future studies could explore this research using an embodied or humanoid robot instead of a virtual robot avatar to validate the research questions.

 The previous studies with proactive robots have generally indicated positive impacts on various aspects of HRI, including trust \cite{Zhu2020Effects}, perceived intelligence \cite{LeMasu2024Reactive}, user experience \cite{chien2022theeffect}, and task performance \cite{Pandey2013Towards}. However, our preliminary results hint towards less acceptance of the disclosure appropriateness of private information in robot's proactive condition compared to the active on-demand condition. The proactive communication of robots offers certain benefits, such as intelligence and trust, but its effect on information disclosure may necessitate careful consideration.







\bibliographystyle{IEEEtran}
\bibliography{myfile}

\end{document}